\documentclass[9pt,conference]{IEEEtran}
\usepackage{amssymb,amsthm,amsmath,array}
\usepackage{graphicx}
\usepackage[caption=false,font=footnotesize]{subfig}
\usepackage{xspace}
\usepackage[sort&compress, numbers]{natbib}
\usepackage{stmaryrd}
\usepackage{xcolor}
\usepackage{mathtools}
\usepackage{float}
\usepackage{textcomp}
\usepackage{float}         %
\usepackage{tikz}      %
\usetikzlibrary{positioning}  %
\usepackage{booktabs}      %
\usepackage{hyperref}
\hypersetup{
    colorlinks,
    linkcolor={black!50!black},
    citecolor={black!50!black},
    urlcolor={black!80!black}
}

\usepackage{amsmath}
\usepackage{algorithm}
\usepackage{algpseudocode}
\usepackage{multicol,multirow}

\newcommand{\C}{\ensuremath{\mathbb{C}}}

\newcommand{\s}{\mathbf{s}}

\newcommand{\A}{\mathbf{A}}

\newcommand{\n}{\mathbf{w}}
\newcommand{\y}{\mathbf{y}}
\newcommand{\x}{\mathbf{r}}

\DeclareMathSymbol{\mathbbE}{\mathord}{AMSb}{"45}

\begin{document}
\title{ Fast Regularized 3D Near-Field MIMO Imaging \\Using Stochastic Proximal Gradient Method}
\author{\IEEEauthorblockN{
        Okyanus Oral
    }
    \IEEEauthorblockA{Interdisciplinary Centre for Security, Reliability and Trust (SnT), University of Luxembourg, L-1855, Luxembourg \\
        Email: okyanus.oral@uni.lu}
}
\maketitle
\begin{abstract} Near-field multiple-input multiple-output (MIMO) radar imaging suffers from high computational load inherently due to irregular spatial sampling with distributed antennas. Existing acceleration methods for near-field MIMO imaging typically rely on interpolation or compensation of measurements and are primarily developed for direct reconstruction. This hinders their ease of adoption for different MIMO geometries and requires further modification for regularized inversion. In this study, we address these challenges by developing a fast regularized reconstruction approach for three-dimensional near-field MIMO imaging based on the Stochastic Proximal Gradient Method. We demonstrate the performance of the developed approach through experimental measurements. The results show a significant improvement in runtime without any notable compromise in reconstruction quality. 
\end{abstract}

\section{Introduction}
For high-resolution radar imaging, multiple-input multiple-output (MIMO) arrays offer low hardware complexity compared to their monostatic counterparts \cite{ahmed2012advanced,anadol2018uwb,yanik2019near,Smith2022Efficient,Wang2020Short-Range}. However, non-uniform or irregularly sampled antenna positions of MIMO arrays limit their use for applications such as near-field radar imaging, in which fast Fourier transform based computationally efficient imaging methods cannot be readily used during image formation \cite{Wang2020Short-Range}. This difference is magnified for iterative regularized reconstruction, especially when applied to large-scale problems such as three-dimensional (3D) imaging \cite{oral2024plug,manisali2024efficient}.  

Most of the approaches to accelerate near-field MIMO imaging consider direct inversion schemes \cite{Wang2020Short-Range,Zhuge2012RMA,yanik2019near,Smith2022Efficient} and are developed as interpolation and/or compensation methods acting on measurements. As a result, these methods require additional care for the geometry of the utilized MIMO array, which reduces their widespread applicability.

In this study, drawing from the stochastic optimization method presented in \cite{Sun2019OnlineTCI} we develop a fast regularized reconstruction method that can be easily utilized irrespective of the choice of array geometry, which to the best of our knowledge is underexplored \cite{Mansour2016Online,Mansour2016Multipath} in the context of near-field MIMO imaging.

\section{Observation Model}

Under Born approximation, we can express the discrete forward model that relates the near-field MIMO radar measurements, $\y~\in~\C^M$, to the discretized reflectivity distribution of the scene, $\s\in\C^N$, in matrix-vector form as follows \cite{Zhuge2012RMA,oral2024plug,manisali2024efficient}:
\begin{equation}
    \y = \A\s + \n.
    \label{eqn:forward-model-MVprod}
\end{equation}
Here $\n\in \C^M$ represents additive uncorrelated Gaussian noise which commonly holds for the practical applications of interest. The matrix $\A \in \C^{M\times N}$ is the observation matrix whose $(m, n)$th element is given by 
\begin{align}
   \A_{m,n}=p(f_m) \frac{\exp({-j\frac{2\pi}{c}f_m\left(\|\x_{T_m}-\x_n\|_2+\|\x_{R_m}-\x_n\|_2\right)})}{4\pi\;\|\x_{T_m}-\x_n\|_2\;\|\x_{R_m}-\x_n\|_2}\, \;,
    \label{eqn:MatrixElements}
\end{align}
and represents the contribution of the $n$th voxel, $\s_n$, at location $\x_n~=~[x_n,y_n,z_n]^T$ to the $m$th measurement, $\y_m$, taken using the transmitter at $\x_{T_m}~=~[x_{Tm},y_{Tm},0]^T$ and the receiver at $\x_{R_m}~=~[x_{Rm},y_{Rm},0]^T$. The temporal Fourier transform of the transmitted pulse is denoted by $p(\cdot)$ with $f_m$ denoting the operating frequency corresponding to the $m$th measurement, and $c$ denoting the speed of light.

\section{Inverse Problem}
In the inverse problem, the goal is to infer the 3D complex-valued reflectivity distribution of the scene, $\s$, from the compressed MIMO radar measurements, $\y$, ($N \gg M$). A common approach is to formulate this inverse problem as an optimization problem of the form 
\begin{equation}\widehat{\s} = \arg \min_{\s} \{\mathcal{D}(\s) + \lambda \mathcal{R}(\s)\},
\label{eq:InverseProblem}
\end{equation}
where the $\mathcal{D}(\cdot)$ and $\mathcal{R}(\cdot)$ respectively represent the data-fidelity and regularization functions; then, the parameter $\lambda > 0$ balances the fidelity of reconstructions to the measurements and the assumed prior information \cite{hansen2010discrete,oktem2019sparsity,Sun2019OnlineTCI,Kamilov2023PhysicsBased,Potter2010}. 

In this study, we use  $\ell_1$ regularization due to its prevalence in radar imaging applications involving scene-reflectivity values with random phases \cite{oral2024plug,cetin2001feature,guven2016augmented,miran2021sparse,Li2015NFCsensing,Potter2010,wang2021rmist}. Nonetheless, the results can be easily extended to other regularization functionals \cite{oral2024plug}. 
Lastly, by considering the linear observation model in \eqref{eqn:forward-model-MVprod} we choose the data fidelity term as mean squared error:
\begin{equation}
    \mathcal{D}(\s)=\frac{1}{2M}\|\y-\A\s\|^2_2
    \label{eq:DataFidelity}
\end{equation}
which can be expressed as the average over component functions, $\mathcal{D}_m(\s) \triangleq \frac{1}{2}|\y_m- \sum^N_{n=1} \A_{m,n}\s_n|^2$ \cite{Sun2019OnlineTCI}.

\section{Regularized Iterative Reconstruction}

As a first-order method, the Proximal Gradient Method (PGM) avoids the computationally intensive matrix inversion typically required to solve normal equations in the data-fidelity updates of second-order optimization methods. This makes it well-suited for 3D near-field MIMO radar~imaging. 

To solve this regularized inverse problem, PGM performs the following fixed-point iterations \cite{Sun2019OnlineTCI,Kamilov2023PhysicsBased}
\begin{equation}
    \s^{(k+1)} = \text{prox}_{\alpha \mathcal{R}}( \s^{(k)} - \eta\nabla_\s \mathcal{D}(\s^{(k)})),
    \label{eq:PGM}
\end{equation}
where $\text{prox}_{\alpha \mathcal{R}}(\cdot)$ is the proximal operator associated with the regularization function and the parameter $\alpha=\lambda\eta$ determines the amount of regularization per iteration. The superscript $k$ denotes the iteration count, and the parameter $\eta>0$ denotes the step size for the gradient descent updates.

\subsection{Fast Regularized Reconstruction via Stochastic PGM}
We use a Stochastic PGM (SPGM) \cite{Sun2019OnlineTCI} to reduce the per-iteration complexity and the runtime of PGM applied to 3D near-field MIMO imaging. 
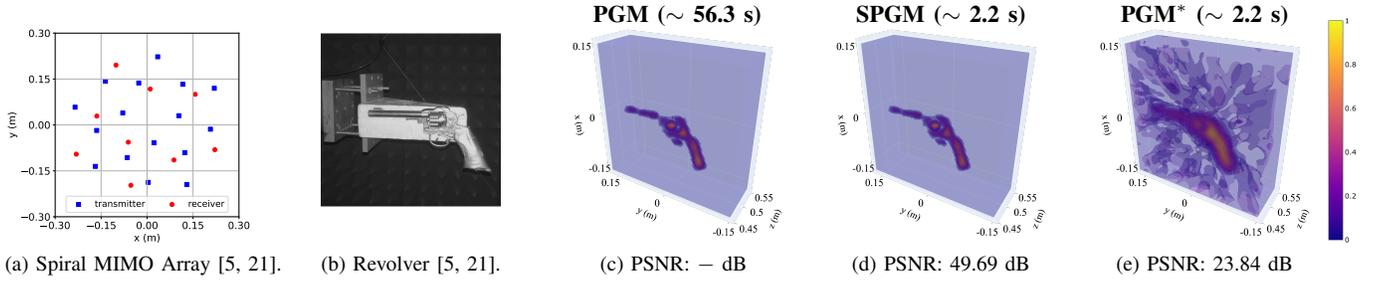
\begin{figure*}[!t]
    \subfloat[Spiral MIMO Array \cite{Wang2020Short-Range,Wang2020EMData}.\label{fig:YarovoyArray}]
    {    
        \includegraphics[width=0.21\linewidth]{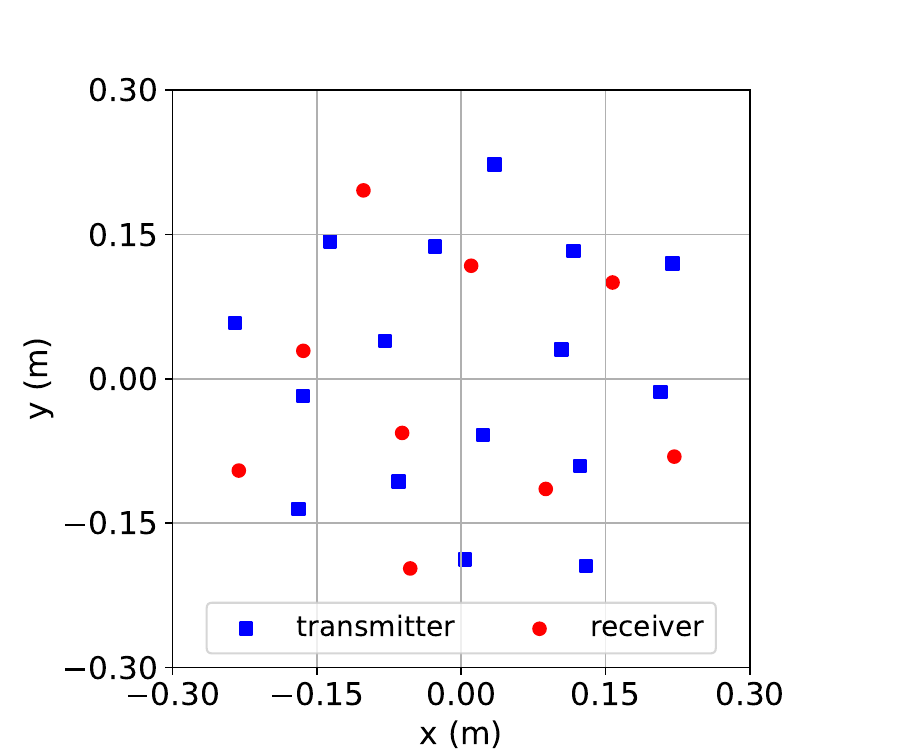}
    }
    \hfill    
    \subfloat[Revolver~\cite{Wang2020Short-Range,Wang2020EMData}.\label{fig:ToyRevolver}]
    {
        \includegraphics[width=0.131\linewidth]{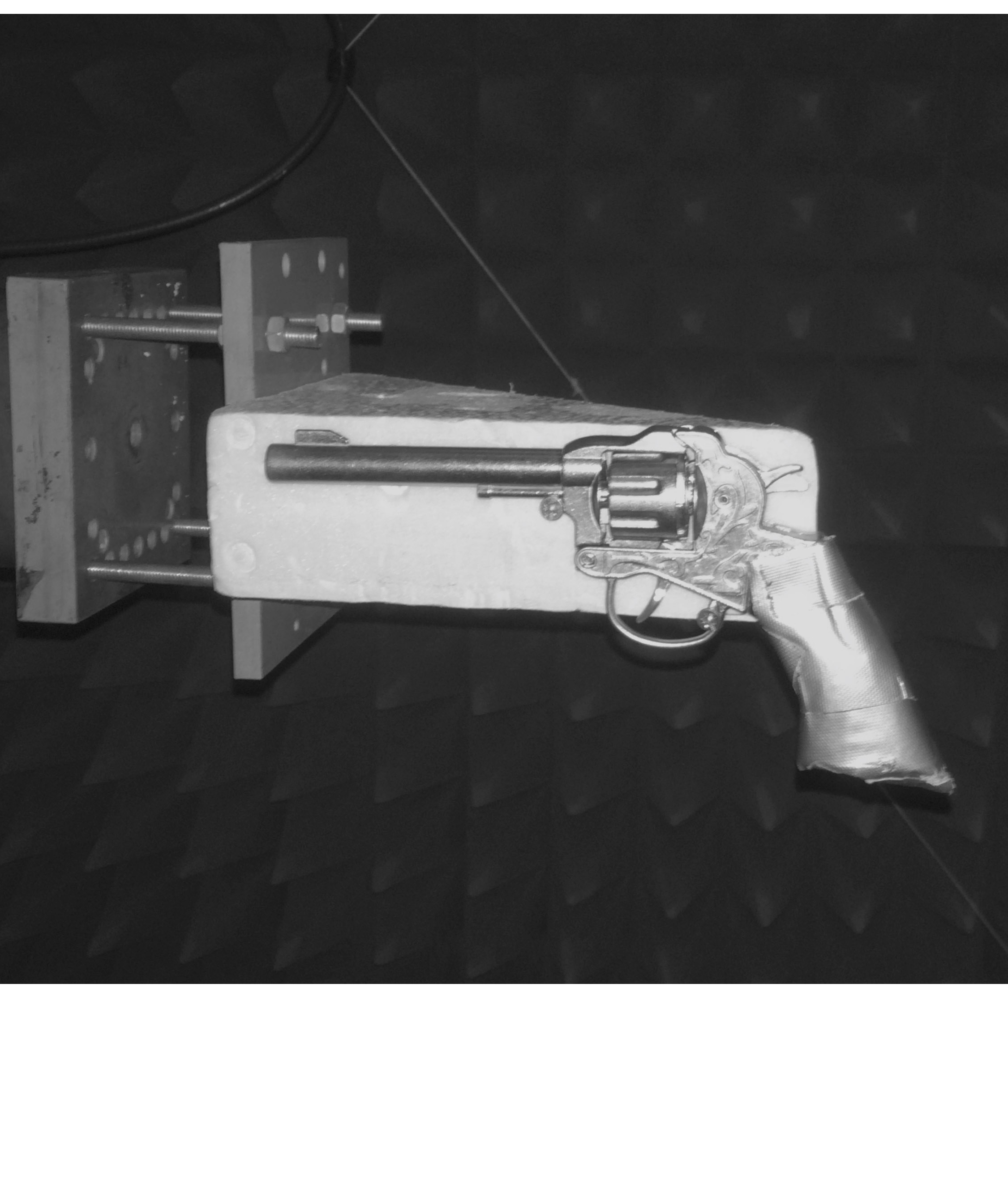}
    }
        \hspace{0.5cm}
    \subfloat[PSNR: $-$ dB \label{fig:PGMreconst}]{\begin{tikzpicture}
            \node[anchor=south west, inner sep=0] (image) at (0,0) {
                \includegraphics[width=0.18\linewidth]{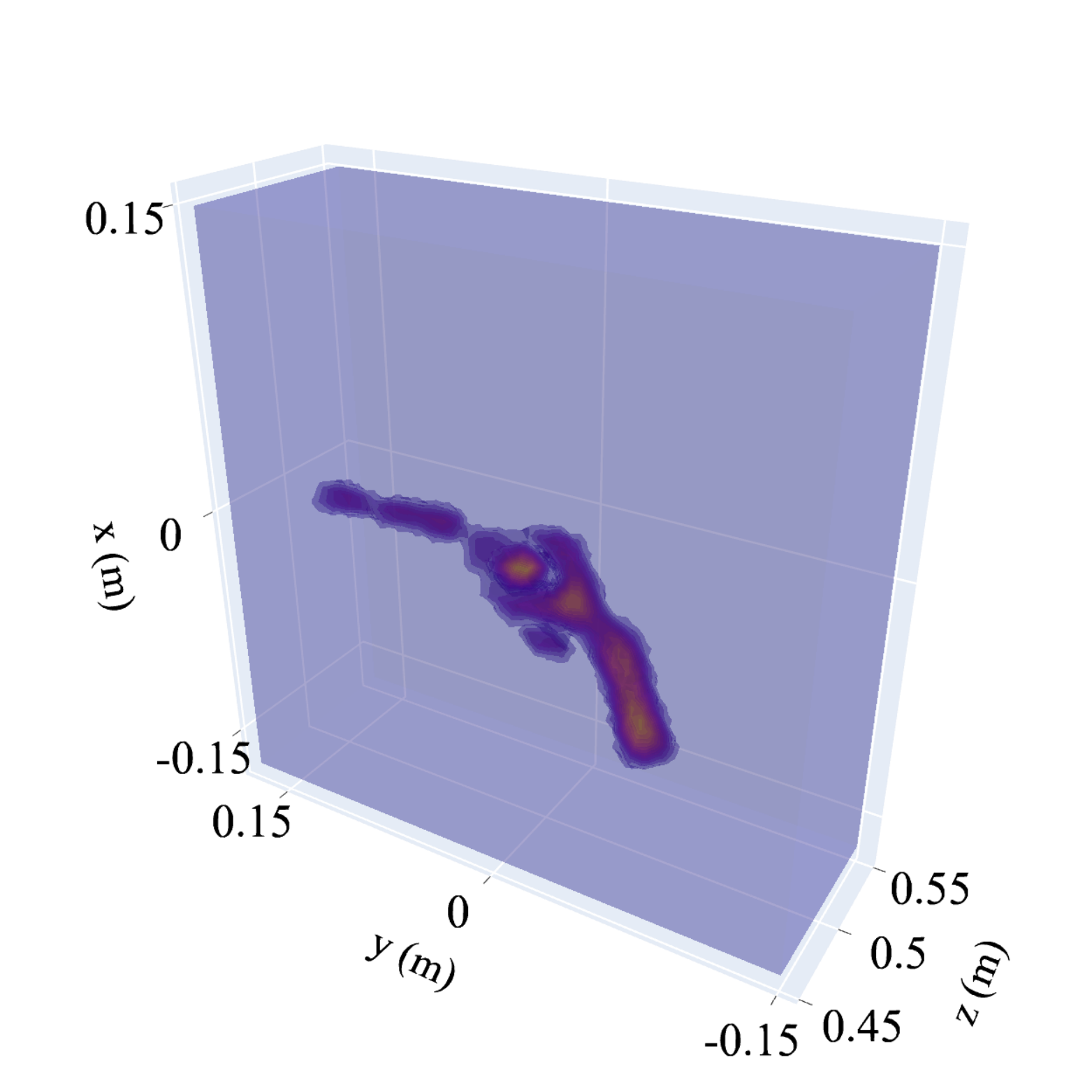}
            };
            \node[above=-0.5cm of image.north, text centered] {\textbf{PGM ($\sim$ 56.3 s)}};  
        \end{tikzpicture}}
    \hfill
    \subfloat[PSNR: 49.69 dB\label{fig:SPGMreconst}]{\begin{tikzpicture}
            \node[anchor=south west, inner sep=0] (image) at (0,0) {
                \includegraphics[width=0.18\linewidth]{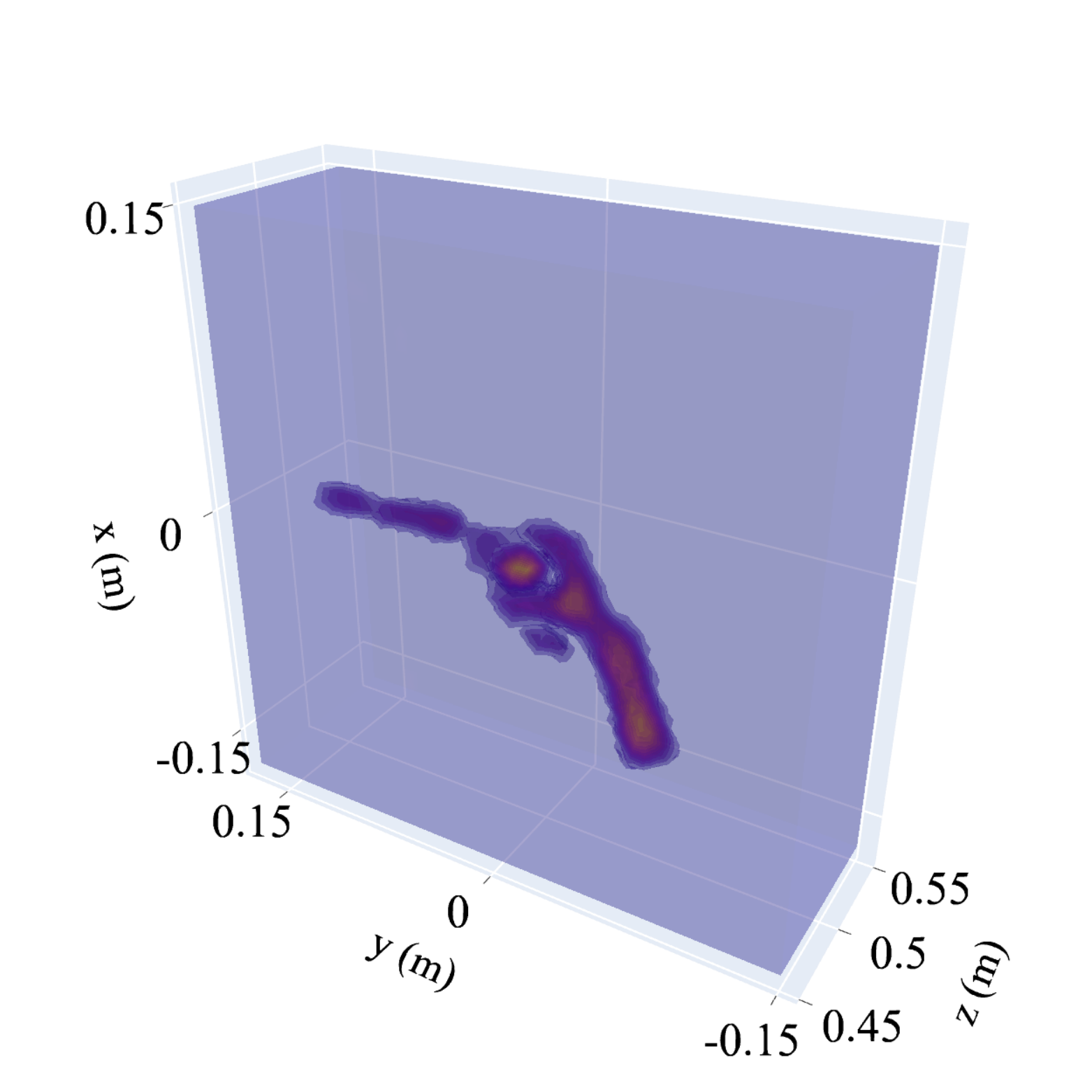}
            };
                  \node[above=-0.5cm of image.north, text centered] {\textbf{SPGM ($\sim$  2.2 s)}};  
        \end{tikzpicture}}
    \hfill
    \subfloat[PSNR: 23.84 dB\label{fig:PGM*reconst}]{\begin{tikzpicture}
            \node[anchor=south west, inner sep=0] (image) at (0,0) {
                \includegraphics[width=0.18\linewidth]{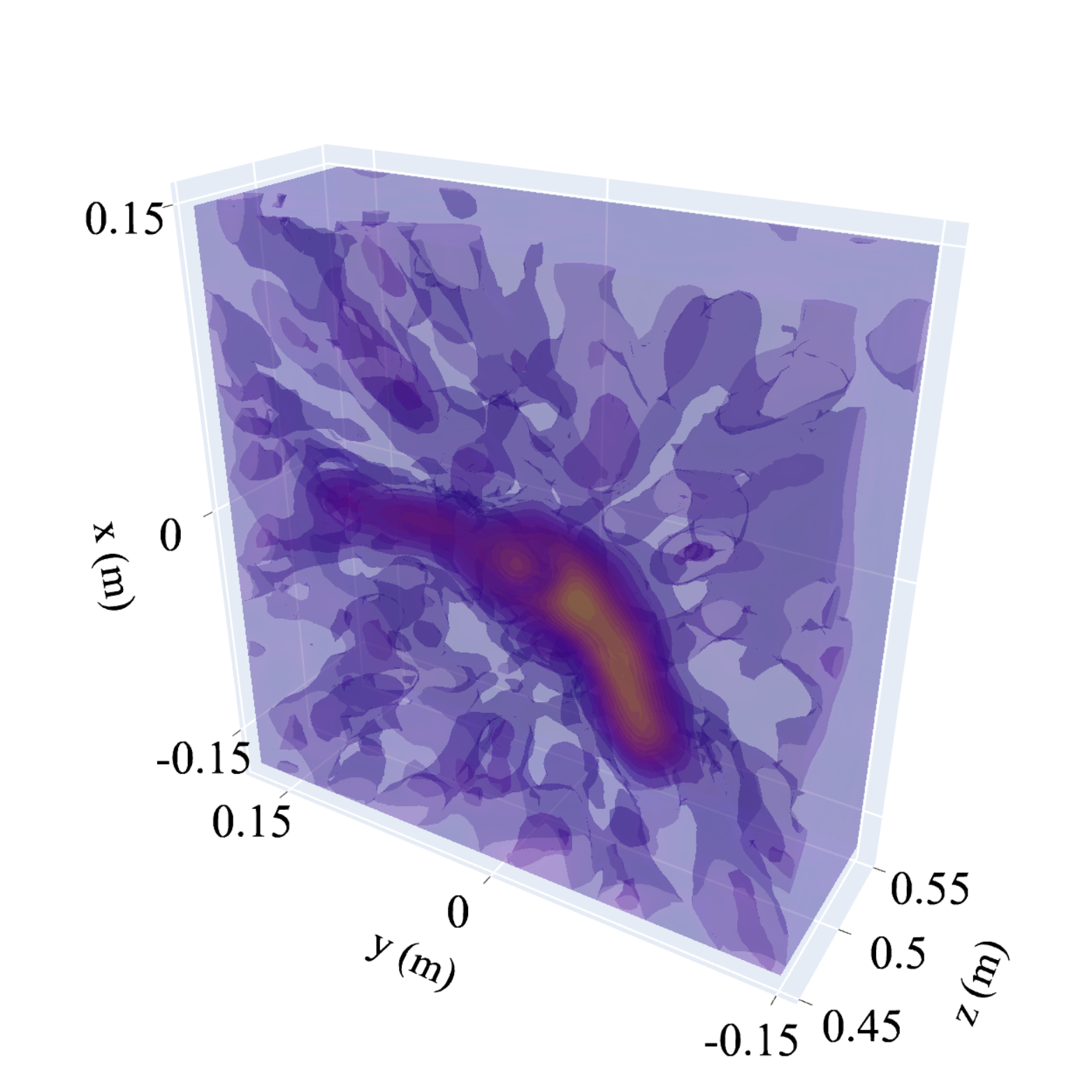}
            };
                        \node[above=-0.5cm of image.north, text centered] {\textbf{PGM$^*$ ($\sim$  2.2 s)}};  
        \end{tikzpicture}}
        \includegraphics[height=3cm]{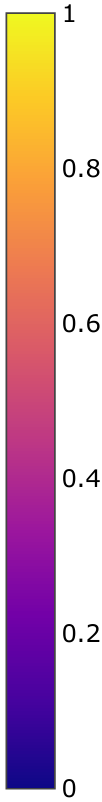}

\caption{ Schematic of the MIMO array corresponding to the experimental measurements \cite{Wang2020Short-Range,Wang2020EMData}, imaged toy revolver \cite{Wang2020Short-Range,Wang2020EMData} and its reconstructions using (c) PGM, (d) SPGM with minibatch of $(\#f,\#T_x,\#R_x)~=~(4,4,3)$, and (d) PGM that is run with the same time-budget as SPGM (denoted as PGM$^*$). PSNR (dB) values of SPGM and fixed budget PGM are indicated beneath the figures and computed using PGM reconstruction as the reference.}
\label{fig:main} 
\end{figure*} 
The main idea of SPGM is to consider \eqref{eq:DataFidelity} as an expectation of the component functions, $\mathcal{D}_m(\s)$, taken over a uniformly distributed random variable $b\in \{1,\dots , M\}$ and to approximate the gradient at every iteration in \eqref{eq:PGM} with an average of $B\ll M$ component gradients \cite{Sun2019OnlineTCI} . 
\begin{equation}
\widehat{\nabla}_\s \mathcal{D}(\s^{(k)}) = \nabla_\s\left( \frac{1}{B} \sum_{m=1}^{B}\mathcal{D}_{b_m}(\s^{(k)})\right)
\end{equation}
Here the parameter $B\geq 1$ is the minibatch size and equals to the product of the number of frequency steps, $\#f$, the number of transmit antennas $\#T_x$, and the number of receive antennas $\#R_x$ used per iteration.

\section{Experiments and Discussions}
We now demonstrate the effectiveness of SPGM applied to 3D near-field MIMO imaging by using experimental measurements available online \cite{Wang2020Short-Range,Wang2020EMData}. These measurements correspond to the MIMO array shown in Fig.\ref{fig:YarovoyArray} that contains 16 transmit and 9 receive antennas; and are associated with the toy revolver shown in Fig.\ref{fig:ToyRevolver} which was placed approximately 50 cm away from the array \cite{Wang2020Short-Range,Wang2020EMData}. Following the scenario in \cite{oral2024plug}, we use 11 equally spaced frequencies from the 4–16 GHz band and infer the reflectivity distribution of a 30cm$\times$30cm$\times$10cm scene using a sampling interval of 0.5 cm. This results in an unknown image cube of 61$\times$61$\times$21 voxels in $x$, $y$, and~$z$ dimensions respectively.

For all of our experiments, we use the same hyperparameters, $\eta=10^{-3}$ and $\alpha=4\cdot10^{-5}$, that are empirically found by qualitatively evaluating PGM reconstructions. The termination criterion is set as the change in reflectivity magnitudes dropping below $10^{-3}$.

\begin{figure}[ht]  
    \centering
     \includegraphics[width=.9\linewidth]{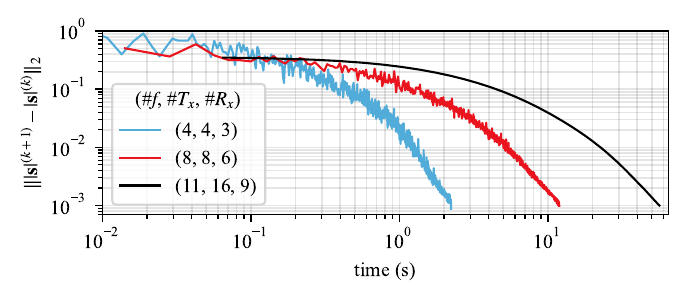}
    \caption{%
    Runtime (s)
    of SPGM. PGM result is shown as using the minibatch of $(\#f,\#T_x,\#R_x)=(11,16,9)$.}
    \label{fig:convplot}
\end{figure}

We first analyze the effect of minibatch size on the runtime.
\begin{figure}[h]
    \centering
    \includegraphics[width=1\linewidth]{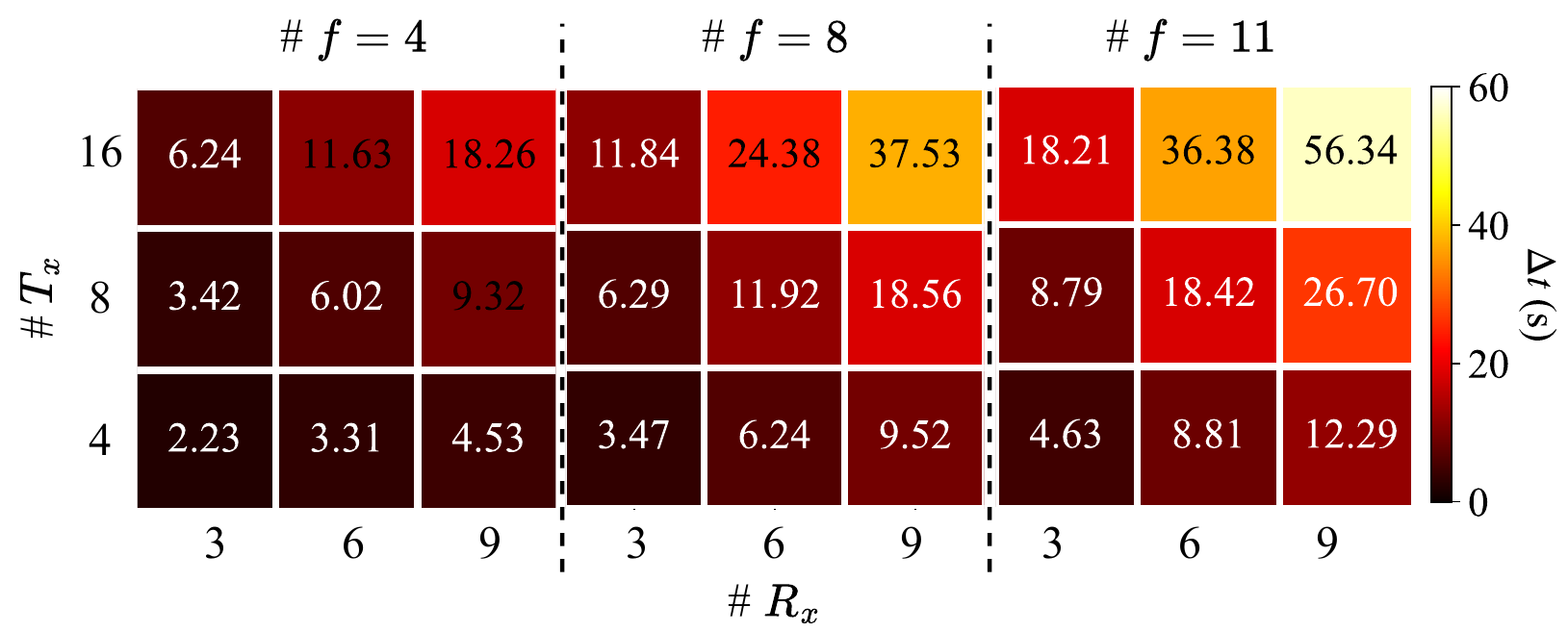}
    \caption{The effect of minibatch composition $(\#f,\#T_x,\#R_x)$ on the runtime (s)
    of SPGM. PGM result is given under the minibatch of $(\#f,\#T_x,\#R_x)=(11,16,9)$.}
    \label{fig:dt}
\end{figure}
As seen, from Fig. \ref{fig:dt} for different minibatch compositions $(\#f,\#T_x,\#R_x)$, the runtime of SPGM is faster for smaller minibatch sizes. 
Most importantly we observe a nearly 25-fold faster %
runtime
compared to PGM when $(\#f,\#T_x,\#R_x)=(4,4,3)$. 
Runtime 
plots for three minibatch compositions are given in Fig.\ref{fig:convplot} for completeness. 

Next, we analyze the accuracy of the fixed-point convergence of SPGM. For this, we compute the peak signal-to-noise ratio (PSNR) over the normalized reflectivity magnitudes \cite{oral2024plug,manisali2024efficient} by taking the PGM reconstruction as the reference image.
\begin{figure}[h]
    \centering
    \includegraphics[width=1\linewidth]{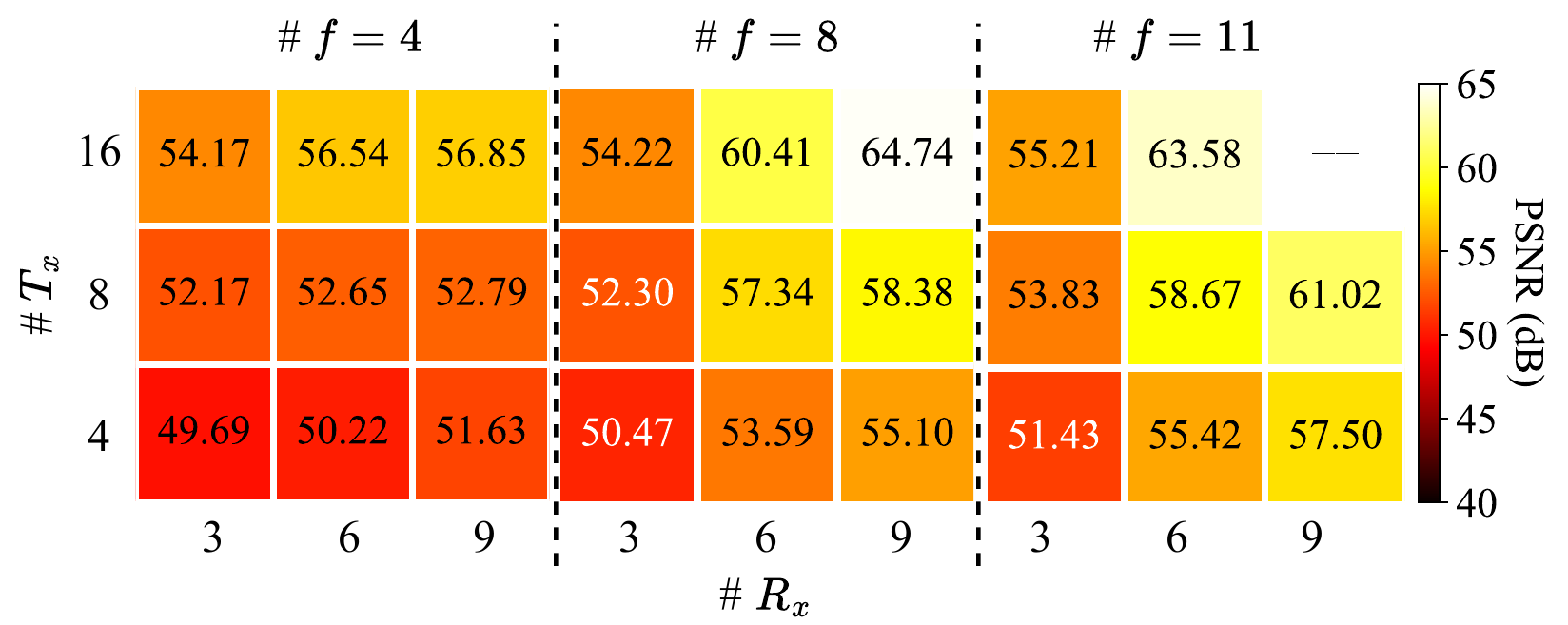}
    \caption{The effect of minibatch composition $(\#f,\#T_x,\#R_x)$ on the accuracy of fixed-point convergence given by PSNR (dB).}
    \label{fig:psnrs}
\end{figure}
As seen from Fig.\ref{fig:psnrs}, although a larger $B$ leads to a higher accuracy, SPGM converges virtually the same set of fixed points as PGM for all of our trials. 

Lastly, the reconstructions of the 3D reflectivity distribution of the scene are shown in Fig.\ref{fig:PGMreconst}-\ref{fig:PGM*reconst}. As seen, SPGM reconstruction is visually indistinguishable from PGM, yet it only takes 2.2 seconds instead of 56.3 seconds. In comparison, when PGM is run with the same time budget as SPGM, we observe severe artifacts. These results, demonstrate the effectiveness of the developed approach for applications requiring both fast computation and high image quality.

\section*{Acknowledgement}
The work is supported by the Luxembourg National Research Fund (FNR) through the CORE project METSA under grant C22/IS/17391632.

\bibliographystyle{IEEEtran}
\bibliography{references}
\end{document}